\documentclass[runningheads]{llncs}
\usepackage{graphicx}
\usepackage{multirow}
\usepackage[colorlinks,linkcolor=red, anchorcolor=blue, citecolor=blue, urlcolor=magenta]{hyperref}
\usepackage{framed,multirow}
\usepackage{threeparttable}
\usepackage[misc]{ifsym}
\usepackage[symbol]{footmisc}
\usepackage{amsmath}
\usepackage{amssymb}
\begin{document}

\title{CoactSeg: Learning from Heterogeneous Data for New Multiple Sclerosis Lesion Segmentation}
\titlerunning{CoactSeg for New MS Lesion Segmentation}

\author{Yicheng Wu\textsuperscript{1(\Letter)} \and Zhonghua Wu \inst{2} \and Hengcan Shi \inst{1} \and Bjoern Picker \inst{3,4} \and Winston Chong \inst{3,4} \and Jianfei Cai\inst{1}}
\authorrunning{Yicheng Wu et al.}

\institute{\textsuperscript{1} Department of Data Science \& AI, Faculty of Information Technology, Monash University, Melbourne, VIC 3168, Australia \\
\email{yicheng.wu@monash.edu}\\
\textsuperscript{2} SenseTime Research, Singapore, 069547, Singapore \\
\textsuperscript{3} Alfred Health Radiology, Alfred Health, Melbourne, VIC 3004, Australia \\
\textsuperscript{4} Central Clinical School, Faculty of Medicine, Nursing and Health Sciences, Monash University, Melbourne, VIC 3800, Australia \\
}
\maketitle 
\begin{abstract}
New lesion segmentation is essential to estimate the disease progression and therapeutic effects during multiple sclerosis (MS) clinical treatments. However, the expensive data acquisition and expert annotation restrict the feasibility of applying large-scale deep learning models. Since single-time-point samples with all-lesion labels are relatively easy to collect, exploiting them to train deep models is highly desirable to improve new lesion segmentation.
Therefore, we proposed a \textbf{coact}ion \textbf{seg}mentation (CoactSeg) framework to exploit the heterogeneous data (\textit{i.e.,} new-lesion annotated two-time-point data and all-lesion annotated single-time-point data) for new MS lesion segmentation. 
The CoactSeg model is designed as a unified model, with the same three inputs (the baseline, follow-up, and their longitudinal brain differences) and the same three outputs (the corresponding all-lesion and new-lesion predictions), no matter which type of heterogeneous data is being used.
Moreover, a simple and effective relation regularization is proposed to ensure the longitudinal relations among the three outputs to improve the model learning.
Extensive experiments demonstrate that utilizing the heterogeneous data and the proposed longitudinal relation constraint can significantly improve the performance for both new-lesion and all-lesion segmentation tasks.
Meanwhile, we also introduce an in-house MS-23v1 dataset, including 38 Oceania single-time-point samples with all-lesion labels. Codes and the dataset are released at \url{https://github.com/ycwu1997/CoactSeg}.
\end{abstract}

\section{Introduction}
Multiple sclerosis (MS) is a common inflammatory disease in the central nervous system (CNS), affecting millions of people worldwide \cite{gold2012placebo} and even leading to the disability of young population \cite{sharmin2022confirmed}. During the clinical treatment of MS, lesion changes, especially the emergence of new lesions, are crucial criteria for estimating the effects of given anti-inflammatory disease-modifying drugs \cite{commowick2021msseg}. However, MS lesions are usually small, numerous, and appear similar to Gliosis or other types of brain lesions, \textit{e.g.,} ischemic vasculopathy \cite{he2022ms}. Identifying MS lesion changes from multi-time-point data is still a heavy burden for clinicians. Therefore, automatically quantifying MS lesion changes is essential in constructing a computer-aided diagnosis (CAD) system for clinical applications.

Deep learning has been widely used for MS lesion segmentation from brain MRI sequences \cite{zeng2020review,tang2021lg}. For example, the icobrain 5.1 framework \cite{rakic2021icobrain} combined supervised and unsupervised approaches and designed manual rules to fuse the final segmentation results. Some works \cite{la2020multiple,zhang2021all} further studied the complementary features from other MRI modalities for MS lesion segmentation. Meanwhile, to train a better deep model, class-imbalance issues \cite{zhang2022qsmrim} and prior brain structures \cite{zhang2023spatially} have been respectively investigated to improve the performance.
With the impressive performance achieved by existing pure MS lesion segmentation methods~\cite{ma2022multiple}, recent attention has been shifted to analyze the longitudinal MS changes \cite{gessert4d,gessert2020multiple}, such as stable, new, shrinking, and enlarging lesions, with the focus on new MS lesion segmentation \cite{kruger2020fully}.

\begin{figure*}[t]
\centering
\includegraphics[width=1.0\textwidth]{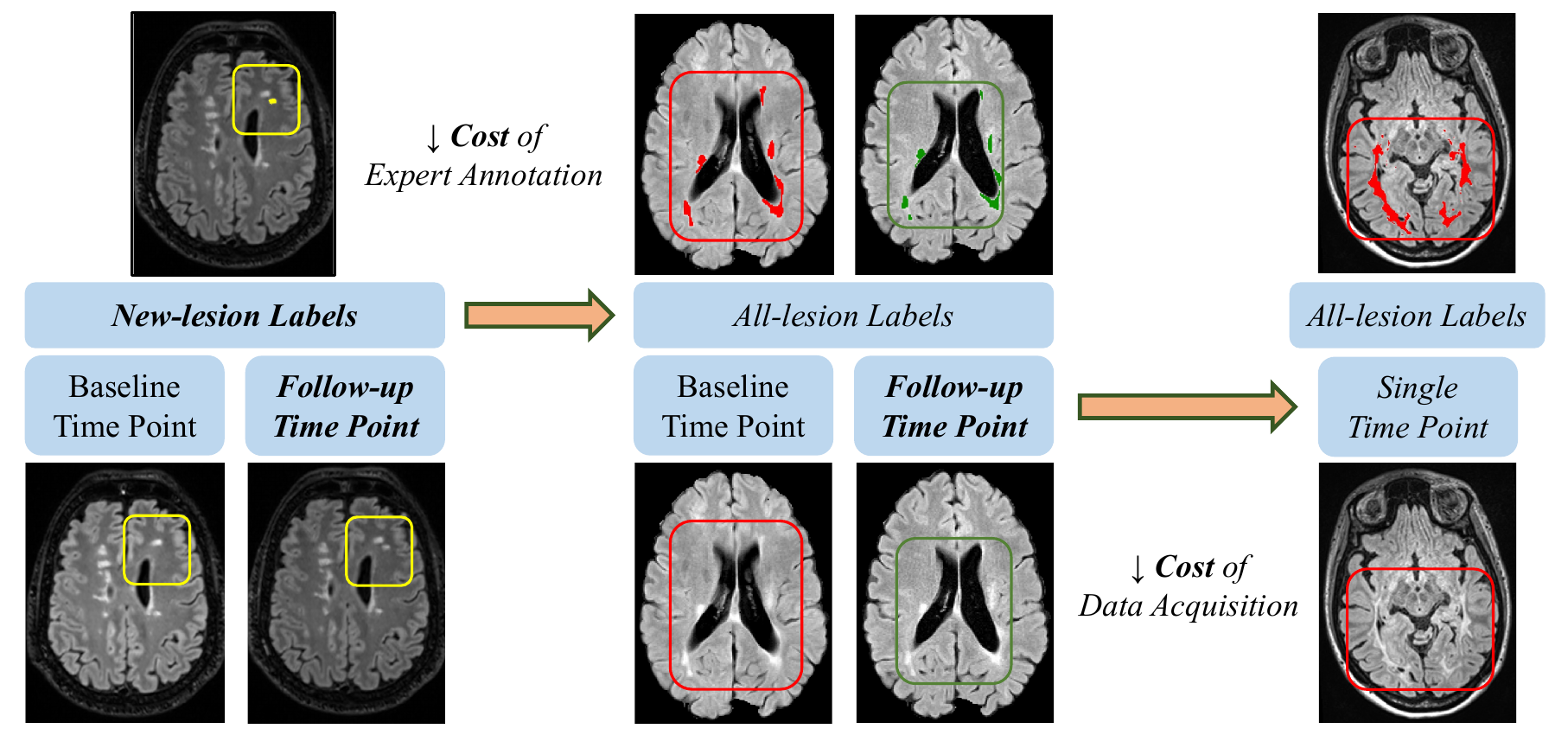}
\caption{\label{motivation}  Heterogeneous data and annotations:  \textit{new-lesion annotated two-time-point data (Left), all-lesion two-time-point data (Middle), and all-lesion single-time-point data (Right)}, with different expert annotation and data acquisition costs. Here, we exploit additional single-time-point data (Right) to help new MS lesion segmentation.}
\end{figure*}

However, collecting adequate well-labeled longitudinal MS lesion data for model learning is highly challenging since it needs multi-time-point data from the same set of patients, and requires costly and time-consuming expert annotations.
Fig.~\ref{motivation} shows the three types of heterogeneous MS lesion data: new-lesion annotated two-time-point data, all-lesion annotated two-time-point data, and all-lesion annotated single-time-point data, each of which is associated with different costs. New-lesion annotated two-time-point data is the ideal one for learning new lesion segmentation, but with the highest data acquisition and annotation costs. Annotating all lesions in two-time-point data can reduce the annotation cost, but it requires accurate brain registration and rule-based post-processing to identify lesion changes, which cannot avoid noise accumulation and often leads to sub-optimal performance. All-lesion annotated single-time-point data is with the cheapest data acquisition and annotation costs. This motivates us to raise the question: \textit{``Can we leverage all-lesion annotated single-time-point data to promote the new MS lesion segmentation?''}

Therefore, in this paper, we proposed a deep \textbf{Coact}ion \textbf{Seg}mentation (CoactSeg) model that can unify  heterogeneous data and annotations for the new MS lesion segmentation task. Specifically,  CoactSeg takes three-channel inputs, including the baseline, follow-up, and corresponding differential brains, and produces all-lesion and new-lesion segmentation results at the same time.
Moreover, a longitudinal relation constraint (\textit{e.g.,} new lesions should only appear at the follow-up scans) is proposed to regularize the model learning in order to integrate the two tasks (new and all lesion segmentation) and boost each other. Extensive experiments on two MS datasets demonstrate that our proposed CoactSeg model is able to achieve superior performance for both new and all MS lesion segmentation, \textit{e.g.,} obtaining 63.82\% Dice on the public MICCAI-21 dataset \cite{commowick2021multiple} and 72.32\% Dice on our in-house MS-23v1 dataset, respectively. It even outperforms two neuro-radiologists on MICCAI-21.

Overall, the contributions of this work are three-fold:
\begin{itemize}
\item We propose a simple unified model CoactSeg that can be trained on both new-lesion annotated two-time-point data and all-lesion annotated single-time-point data in the same way, with the same input and output format;
\item We design a relation regularizer to ensure the longitudinal relations among all and new lesion predictions of the baseline, follow-up, and corresponding differential brains;
\item We construct an in-house MS-23v1 dataset, which includes 38 Oceania single-time-point 3D FLAIR scans with manual all-lesion annotations by experienced human experts. We will release this dataset publicly.
\end{itemize}

\section{Datasets}
{
\begin{table*}[!htb]
	\centering
	\caption{Details of the experimental datasets in this work. Note that the data split is fixed and shown in the 4th column (Order: Training, Validation).}
	\label{dataset}
    \begin{threeparttable}
	\resizebox{1\textwidth}{!}{
	\begin{tabular}{c|c|c|c|c|c}
		\hline 
		\hline
		Dataset&Region&Modality&\# of subjects & \# of time points & Annotation Type\\
  \hline
  MICCAI-21 \cite{commowick2021multiple}&France& FLAIR&40 (32, 8) & 2 & New lesions\\
    \hline
  MS-23v1 (Ours) &Oceania&FLAIR&38 (30, 8) & 1 & All lesions\\
		\hline
		\hline
	\end{tabular}}
    \end{threeparttable}
\end{table*}
}
We trained and evaluated our CoactSeg model on two MS segmentation datasets, as shown in Table~\ref{dataset}. On the public MICCAI-21 dataset\footnote{\url{https://portal.fli-iam.irisa.fr/msseg-2/}}, we only use its training set since it does not provide official labels of testing samples. Specifically, 40 two-time-point 3D FLAIR scans are captured by 15 MRI scanners at different locations. Among them, 11 scans do not contain any new MS lesions. The follow-up data were obtained around 1-3 years after the first examination. Four neuro-radiologists from different centers manually annotated new MS lesions, and a majority voting strategy was used to obtain the final ground truth. For pre-processing, the organizers only performed a rigid brain registration, and we further normalized all MRI scans to a fixed resolution of [0.5, 0.75, 0.75] mm.

Since the public MS lesion data is not adequate \cite{commowick2021multiple,commowick2018objective,carass2017longitudinal}, we further collected 38 single-time-point 3D FLAIR sequences as a new MS dataset (MS-23v1). Specifically, all samples were anonymized and captured by a 3T Siemens scanner in Alfred Health, Australia. To the best of our knowledge, this will be \textit{the first open-source dataset from Oceania} for MS lesion segmentation, contributing to the diversity of existing public MS data. Two neuro-radiologists and one senior neuro-scientist segmented all MS lesions individually and in consensus using the MRIcron segmentation tool\footnote{\url{https://www.nitrc.org/projects/mricron/}}. The voxel spacing of all samples is then normalized to an isotropic resolution of [0.8, 0.8, 0.8] mm. 

Finally, when conducting the mixed training, we used a fixed data split in this paper (\textit{i.e.,} 62 samples for training and 16 for validation in total). Note that we followed the setting of the public challenge \cite{commowick2021multiple}, which selects the new validation set from MICCAI-21 that does not include samples without any new MS lesions.

\section{Method}

\subsection{Overview}
\begin{figure*}[t]
\centering
\includegraphics[width=1.0\textwidth]{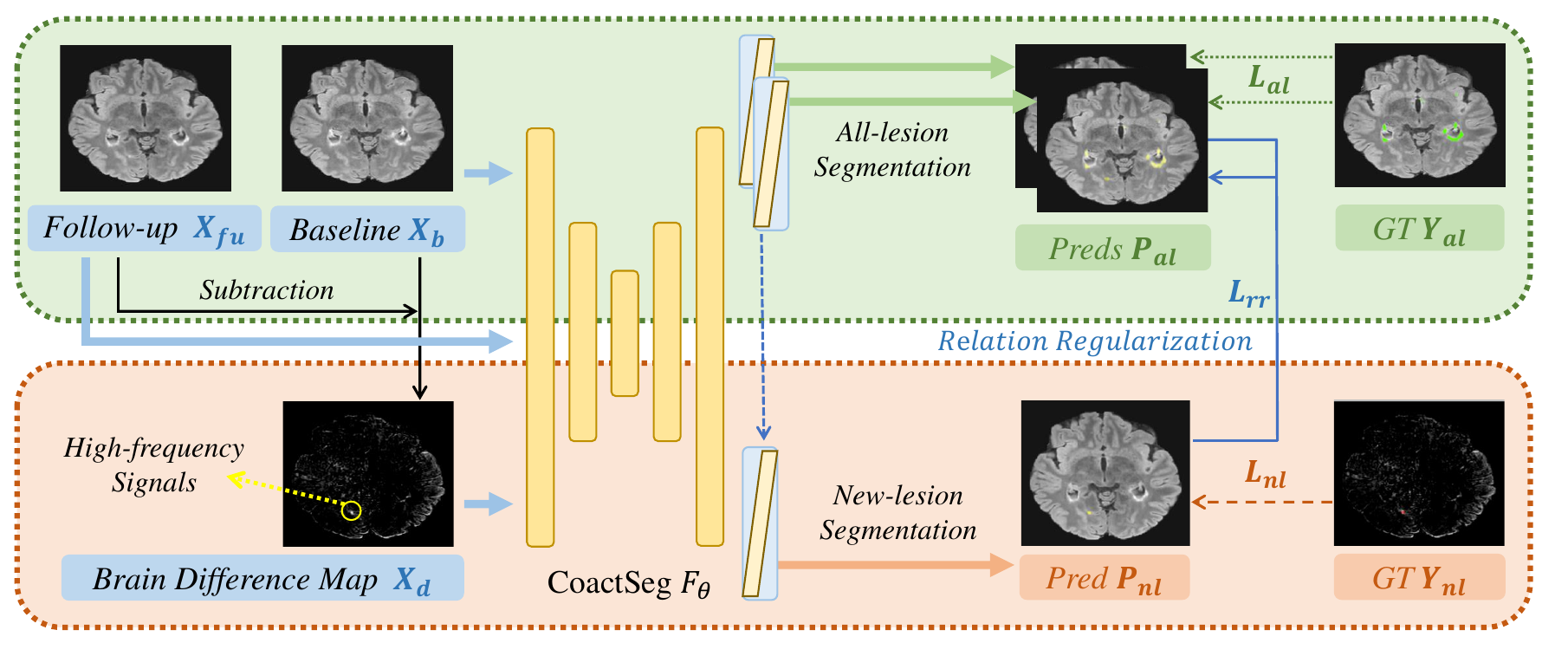}
\caption{\label{pipeline} Pipeline of our proposed CoactSeg model $F_\theta$, which receives the baseline $X_b$, follow-up $X_{fu}$, and corresponding longitudinal brain differences $X_d$ as inputs. $F_\theta$ segments all lesions $P_{al}$ and predicts new lesions $P_{nl}$ in the condition of $X_b$, $X_{fu}$ and $X_d$. Note that new-lesion regions would have higher intensities in $X_d$.}
\end{figure*}

Fig.~\ref{pipeline} illustrates the overall pipeline of our proposed CoactSeg model $F_\theta$. We construct a quadruple set ${(X_b, \ X_{fu}, \ X_d, \ Y)}$ for the model training. Here, the longitudinal difference map $x_d \in X_d$ is obtained by a subtraction operation between the baseline brain $x_b \in X_b$ and its follow-up  $x_{fu} \in X_{fu}$ (\textit{i.e.,} $x_d = x_{fu}-x_b$). Therefore, given heterogeneous annotations, \textit{i.e.,} all-lesion labels $y_{al}^s \in Y_{al}^s$ in single-time-point data and new-lesion labels $y_{nl}^t \in Y_{nl}^t$ in two-time-point data, the CoactSeg model $F_\theta$ is designed to exploit both for the model training.

\subsection{Multi-head Architecture}
Fig.~\ref{pipeline} shows that new-lesion regions are highlighted in the brain difference map $x_d$. Hence, besides $x_b$ and $x_{fu}$, CoactSeg also receives $x_d$ as inputs. It  generates all-lesion and new-lesion predictions  as
\begin{equation}
\begin{aligned}
&p_{al}^{s1}, \ p_{al}^{s2}, \ p_{nl}^s =  F_\theta(x_b^s, \ x_{fu}^s, \ x_d^0)  \\
&p_{al}^{t1}, \ p_{al}^{t2}, \ p_{nl}^t =  F_\theta(x_b^{t}, \ x_{fu}^{t}, \ x_d^t).
\end{aligned}
\end{equation}
For single-time-point samples $x^s \in X^s$, $x_b^s$ and $x_{fu}^s$ are identical as $x^s$, and the difference map becomes an all-zero matrix $x_d^0$, with $p_{al}^{s1}$, $p_{al}^{s2}$ and $p_{nl}^{s}$ being the corresponding all-lesion and new-lesion predictions of $x^s$. For two-time-point data $x^t \in X^t$, 
$x_b^{t}$ and $x_{fu}^{t}$ respectively denote the first and second time-point data samples, with $p_{al}^{t1}$,  $p_{al}^{t2}$ and $p_{nl}^{t}$ being the all-lesion segmentation results at the first and second time-point and the new-lesion results of $x^t$, respectively.

In this way, we unify the learning of both single and two-time-point data with heterogeneous annotations by using the same model $F_\theta$, with the same input and output formats.
Note that, inspired by semi-supervised learning~\cite{wu2022mutual,wu2022exploring,mcnet}, we mix $x^s$ and $x^t$ samples into each batch for training. Given the heterogeneous annotations, \textit{i.e.,} all-lesion labels for single-time-point data and new-lesion labels for two-time-point data, we apply the following corresponding supervisions:
\begin{equation}
\begin{aligned}
&L_{al} =  Dice(p_{al}^{s1}, \ y_{al}^s) + Dice(p_{al}^{s2}, \ y_{al}^s)  \\
&L_{nl} =  Dice(p_{nl}^t, \ y_{nl}^t)
\end{aligned}
\end{equation}
where $Dice$ refers to the common Dice loss for medical segmentation tasks. We use a 3D VNet \cite{vnet} as the backbone of $F_\theta$ and three prediction heads are designed as individual convolutional blocks. Note that, the last prediction head also receives the features from the first two in order to capture the all-lesion information. Compared to the recent work \cite{zhang2021dodnet} for exploiting heterogeneous data, our architecture avoids the complicated design of dynamic prediction heads.

\subsection{Longitudinal Relation Regularization}
Human experts usually identify new MS lesions by comparing the brain MRI scans at different time points. Inspired by this, we further propose a longitudinal relation constraint to compare samples from different time points:
\begin{equation} \label{eq:rr}
L_{rr} =  ||p_{al}^{s1}, \ p_{al}^{s2}||_2 + ||p_{al}^{t1} \otimes y_{nl}^t, \ 0||_2 + ||p_{al}^{t2} \otimes y_{nl}^t, 1||_2
\end{equation}
where $\otimes$ is a masking operation. The first term in \eqref{eq:rr} is to encourage the all-lesion predictions $p_{al}^{s1}$ and $p_{al}^{s2}$ to be the same since there is no brain difference for single-time-point data. The second and third terms in \eqref{eq:rr} are to ensure  that the new-lesion region can be correctly segmented as the foreground in $p_{al}^{t2}$ and as the background in $p_{al}^{t1}$ in two-time-point data with only new lesion labels $y_{nl}^t$. 

Finally, the overall loss function to train our CoactSeg model becomes a weighted sum of $L_{al}$, $L_{nl}$, and the regularization $L_{rr}$:
\begin{equation}
L =  L_{al} + \lambda_1 \times L_{nl} +\lambda_2 \times L_{rr} 
\end{equation}
where $\lambda_1$ and $\lambda_2$ are constants to balance different tasks.

\section{Results}
\subsubsection{Implementation Details.}\label{details} For training, we normalize all inputs as zero mean and unit variance. Then, among common augmentation operations, we use the random flip or rotation to perturb inputs. Since MS lesions are always small, we apply a weighted cropping strategy to extract 3D patches of size $80\times80\times80$ to relieve the class imbalance problem \cite{zhang2022qsmrim}. Specifically, if the input sample contains the foreground, we randomly select one of the foreground voxels as the patch center and shift the patch via a maximum margin of [-10, 10] voxels. Otherwise, we randomly crop 3D patches. The batch size is set as eight (\textit{i.e.,} four new-lesion two-time-point samples and four all-lesion single-time-point samples). We apply Adam optimizer with a learning rate of 1e-2. The overall training iterations are 20k. In the first 10k iterations, $\lambda_1$ and $\lambda_2$ are set to 1 and 0, respectively, in order to train the model for segmenting MS lesions at the early training stage. After that, we set $\lambda_2$ as 1 to apply the relation regularization. During testing, we extract the overlapped patches by a stride of $20\times20\times20$ and then re-compose them into the entire results. 

Note that we follow \cite{schell2019automated} to mask the non-brain regions and all experiments are only conducted in the brain regions with the same environment (Hardware: Single NVIDIA Tesla V100 GPU; Software: PyTorch 1.8.0, Python 3.8.10; Random Seed: 1337). The computational complexity of our model is 42.34 GMACs, and the number of parameters is 9.48 M.
{
\begin{table*}[!htb]
	\centering
	\caption{Comparisons of new-lesion segmentation on MICCAI-21. Note that the human experts' performance is shown based on their individually annotated results.}
	\label{comparison}
    \begin{threeparttable}
	\resizebox{0.8\textwidth}{!}{
	\begin{tabular}{c|ccccc}
		\hline 
		\hline
		\multirow{2}{*}{Method}&\multicolumn{5}{|c}{Performance on MICCAI-21 (New MS Lesions)} \\
  \cline{2-6}
& Dice(\%)$\uparrow$ & Jaccard(\%)$\uparrow$ &95HD(voxel)$\downarrow$ &ASD(voxel)$\downarrow$
  &F1(\%)$\uparrow$ \\
    \hline
    SNAC \cite{snac} & 53.07 &39.19&66.57 &26.39&30.71 \\
    SNAC (VNet) \cite{snac} &56.81 &42.85&26.58 &12.49&57.59 \\
    Neuropoly \cite{macar2021team} &56.33 &43.23&54.95 &24.16&17.47 \\
    CoactSeg (Ours) &63.82 &51.68&30.35 &12.14&61.96\\
  \hline
    Human Expert \#1 &77.52 &65.76&27.83 &5.47&82.34 \\
    Human Expert \#2 &66.89 &58.11 &N/A &N/A&68.19 \\
    Human Expert \#3 &58.56 &46.51&60.99 &12.41&62.88 \\
    Human Expert \#4 &60.68&49.95 &N/A &N/A&66.58 \\
    \hline
    \hline
	\end{tabular}}
    \end{threeparttable}
\end{table*}
}
\subsubsection{Performance for MS Lesion Segmentation.} Two MS tasks (\textit{i.e.,} new-lesion segmentation on MICCAI-21 and all-lesion segmentation on our MS-23v1 dataset) are used to evaluate the proposed CoactSeg. Besides common segmentation metrics \cite{maier2022metrics} including Dice, Jaccard, 95\% Hausdorff Distance (95HD), and Average Surface Distance (ASD), we further follow \cite{commowick2018objective} to use the instance-level F1 score (F1) to denote the lesion-wise segmentation performance. Here, tiny lesions (\textit{i.e.,} fewer than 11 voxels) are not included in the F1 calculation as \cite{commowick2018objective}.

\begin{figure*}[t]
\centering
\includegraphics[width=1.0\textwidth]{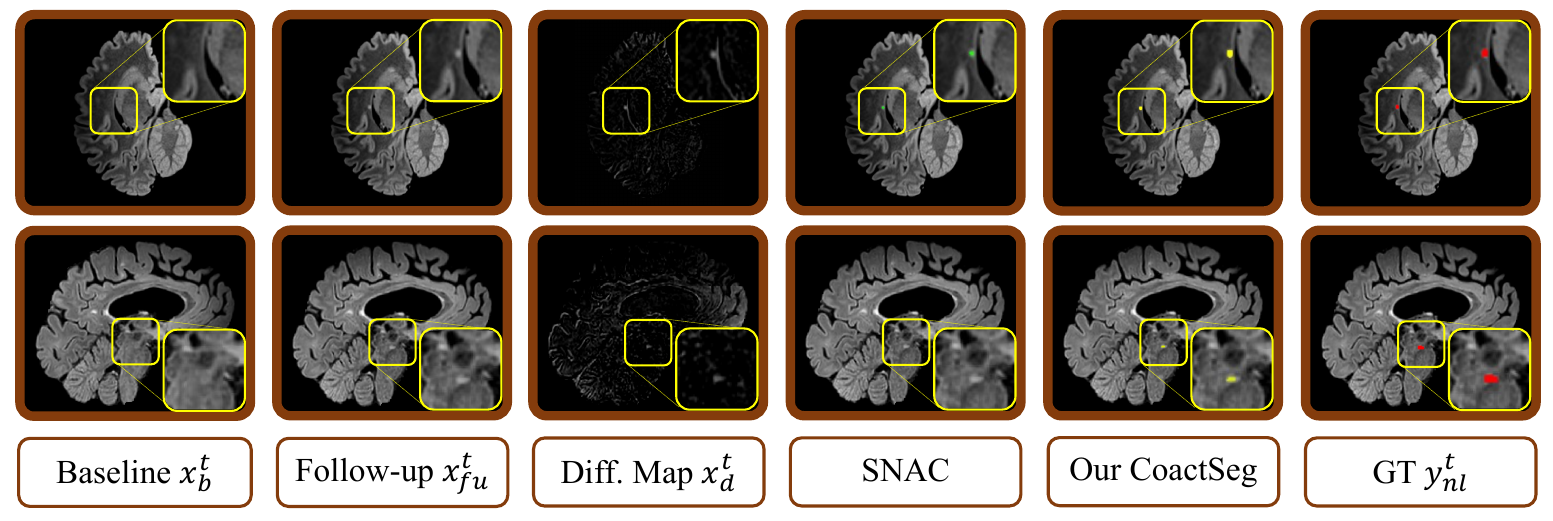}
\caption{\label{new_lesions} Exemplar results for new MS lesion segmentation on the MICCAI-21 dataset.}
\end{figure*}
\begin{figure*}[t]
\centering
\includegraphics[width=1.0\textwidth]{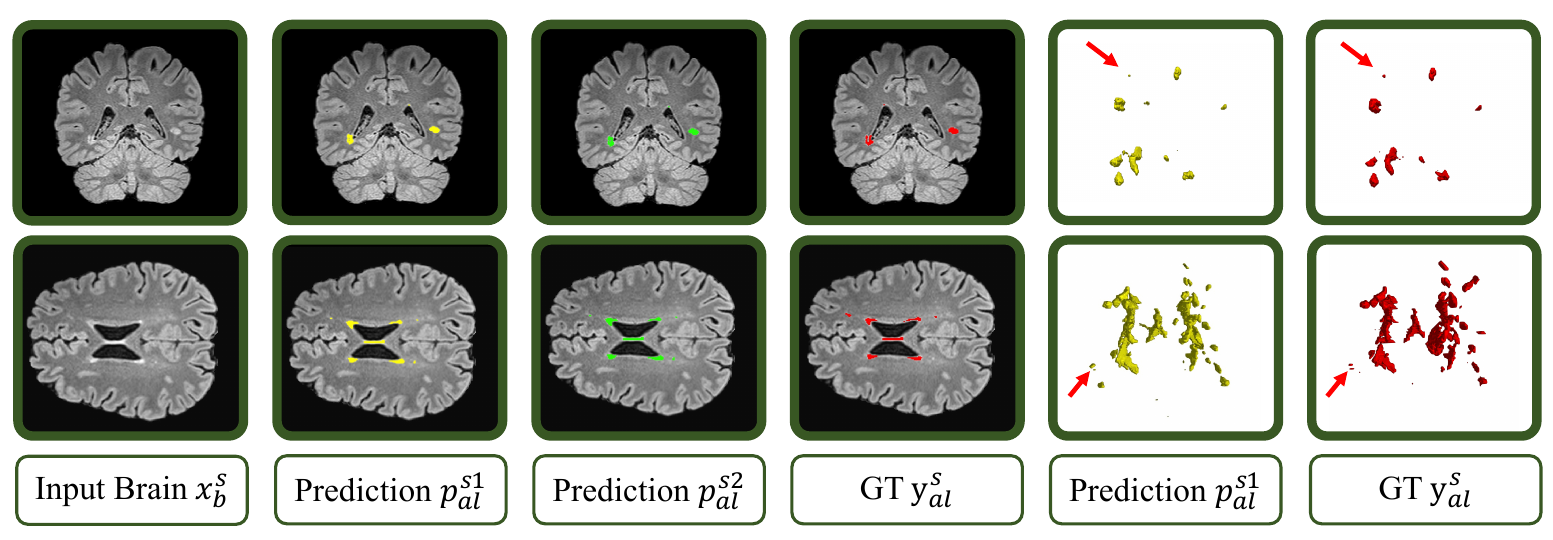}
\caption{\label{all_lesions} Exemplar results for all MS lesion segmentation obtained by our CoactSeg model on our in-house MS-23v1 dataset (2D View: Left; 3D View: Right). }
\end{figure*}

Fig.~\ref{new_lesions} illustrates that our proposed CoactSeg accurately segments the tiny new lesions on MICCAI-21. Compared to the recent work \cite{snac}, our model can even predict new lesions with low contrast (indicated by the enlarged yellow rectangles in Fig.~\ref{new_lesions}). Table~\ref{comparison} gives the quantitative results on MICCAI-21. We can see that: 1) Our model achieves good segmentation performance for new MS lesion segmentation and outperforms the second-best method \cite{snac} by 7.01\% in Dice; 2) Compared with human experts, our proposed model also outperforms two of them (\textit{i.e.,} \#3 and \#4) in terms of the segmentation and the shape-related metrics; 3) For the lesion-wise F1 score, our method 
significantly reduces the performance gap between deep models and human experts, achieving a comparable F1 with expert \#3 (\textit{i.e.,} 61.96\% vs. 62.88\%).

Fig.~\ref{all_lesions} shows the all-lesion segmentation results of our CoactSeg model on our in-house MS-23v1 dataset. It can be seen that CoactSeg is able to segment most MS lesions, even for very tiny ones (highlighted by red arrows). Moreover, we can see that the segmentation results of the first two prediction heads are relatively consistent (\textit{i.e.,} the 2nd and 3rd columns of Fig.~\ref{all_lesions}), demonstrating the effectiveness of our proposed relation regularization.

{
\begin{table*}[!htb]
	\centering
	\caption{Ablation studies of our proposed CoactSeg model $F_\theta$ and $*$ indicates that we apply a stage-by-stage training strategy in the experiments. }
	\label{ablation}
    \begin{threeparttable}
	\resizebox{\textwidth}{!}{
	\begin{tabular}{c|cc|ccc|ccc}
		\hline 
		\hline
		\multirow{2}{*}{$L_{rr}$}&\multicolumn{2}{c}{Training Data}&\multicolumn{3}{|c}{MICCAI-21 (New MS Lesions)}&\multicolumn{3}{|c}{MS-23v1 (All MS Lesions)} \\
  \cline{2-9}
		&MICCAI-21&MS-23v1& Dice(\%)$\uparrow$ &95HD(voxel)$\downarrow$&F1(\%)$\uparrow$&Dice(\%)$\uparrow$ &95HD(voxel)$\downarrow$&F1(\%)$\uparrow$ \\
  \hline
  w/o&\checkmark&  &59.91&35.73&45.61&\multicolumn{3}{c}{N/A} \\
    w/o&&\checkmark  &\multicolumn{3}{c|}{N/A}&70.94  &14.46&\textbf{44.82}\\
    w/o&\checkmark&\checkmark &61.53 &43.05 &51.54 &69.41  &18.93 &34.43 \\
    \hline
  w/&\checkmark& &58.49  &50.17 &51.35&\multicolumn{3}{c}{N/A} \\
    w/&&\checkmark  &\multicolumn{3}{c|}{N/A}&71.28  &12.45 &42.96\\
    w/&\checkmark&\checkmark &62.15  &43.26 &56.97 &70.44 &12.88 &44.04\\
    \hline
    w/&\checkmark*&\checkmark* &\textbf{63.82} & \textbf{30.35} &\textbf{61.96}&\textbf{72.32}  &\textbf{12.38} &42.51\\
    \hline
    \hline
\end{tabular}}
    \end{threeparttable}
\end{table*}
}
\subsubsection{Ablation Study.} Table~\ref{ablation} further shows the ablation study for both new and all MS lesion segmentation tasks. It reveals that: 1) Introducing the heterogeneous data significantly improves the performance of new-lesion segmentation on MICCAI-21 with an average Dice gain of 2.64\%; 2) Exploiting the relation regularization for mixed training can further improve the performance on the two datasets; 3) The simple stage-by-stage training strategy (See the \textit{Implementation Details} \ref{details}) can better balance two tasks and achieve the overall best segmentation performance for both tasks.

\section{Conclusion}
In this paper, we have presented a unified model CoactSeg for new MS lesion segmentation, which can predict new MS lesions according to the two-time-point inputs and their differences while at the same time segmenting all MS lesions. Our model effectively exploits heterogeneous data for training via a  multi-head architecture and a relation regularization. Experimental results demonstrated that introducing all-lesion single-time-point data can significantly improve the new-lesion segmentation performance. Moreover, the relation constraint also facilitates the model to capture the longitudinal MS changes, leading to a further performance gain. Our in-house MS-23v1 dataset will be made public to help the MS lesion research.
Future works will explore more longitudinal relations to study the fine-grained MS changes as well as consider more powerful constraints to address the domain gap \cite{wolleb2022learn} and fairness \cite{zhang2023interaction} problems. Moreover, we plan to collect and annotate more MS lesion data to improve the possibility of training large-scale deep models for clinical applications \cite{reuter2012within}.

\subsubsection{Acknowledgement.} This work was supported in part by the Monash FIT Start-up Grant, in part by the Novartis (ID: 76765455), and in part by the Monash Institute of Medical Engineering (MIME) Project: 2022-13. We here appreciate the public repositories of SNAC \cite{snac} and Neuropoly \cite{macar2021team}, and also thanks for the efforts to collect and share the MS dataset \cite{commowick2021msseg} and the MS-23v1 dataset from Alfred Health, Australia.
\bibliographystyle{splncs04}
\bibliography{paper.bib}
\end{document}